\documentclass[10pt, notitlepage]{article}
\usepackage{calc}
\usepackage{natbib}
\usepackage{amsmath}
\usepackage{amscd}
\usepackage{amssymb}
\usepackage{amsfonts}
\usepackage{amsthm}
\usepackage{amsxtra}
\usepackage{euscript}
\usepackage{changebar}
\usepackage{graphicx}
\usepackage{multicol}
\usepackage{epsfig}
\usepackage{psfrag}

 \title{Summarization and Classification of Non-Poisson Point Processes}
\author{jeffrey Picka and Mingxia Deng \\
Department of Mathematics and Statistics, University of New Brunswick}

\begin{document}

\maketitle

\begin{abstract}
Fitting models for non-Poisson point processes is complicated by the lack of tractable models for much of the data. By using large samples of independent and identically distributed realizations and statistical learning, it is possible to identify absence of fit through finding a classification rule that can efficiently identify single realizations of each type. The method requires a much wider range of descriptive statistics than are currently in use, and a new concept of model fitting which is derive from how physical laws are judged to fit data. 
\end{abstract}

\section{Introduction}

The statistical analysis of point processes is well understood, provided that the data is modeled by a homogeneous or inhomogeneous Poisson point process. If there are interactions among the points, the models become more complicated and so does the inference. In the case of many sets of data, using tests to see if probabilistically defined models fit is very difficult, both in terms of finding models to fit and tests which are sensible. In the absence of either, there is little that can be done using existing theory. Taking a more experimental approach based on statistical learning has the potential to circumvent these difficulties.

For the most part, the non-Poisson point processes will be assumed to be regular point processes, in which no two points can be any closer than some distance $\delta > 0$. The methods developed here could be applied to any point process that defies modeling by a straightforward probabilistic model. 

The methods discussed in this paper could be applied to four different types of problem in point process inference. 

\begin{enumerate}
\item In a pathology study, images of tissue are available in which the nuclei of cells have been stained. The centres of the stained nuclei can be modeled by a regular point pattern, but the pattern is different in healthy tissue and in cancerous tissue. Can a classification rule be found which can identify single samples of tissue as being healthy or cancerous?

\item A formal model for a non-Poisson process is proposed. It is mathematically intractable, but algorithms are proposed which allow realizations to be simulated after some simplifying assumptions are made. There are two such algorithms, built using slightly different simplifying assumptions. Is there any evidence that the two algorithms are sampling from different point processes? 

\item Two plants are constructed in different locations by different constructors to make a cement-fibre composite with fibres aligned in parallel. Cross sections are taken through specimens of the composite from both plants. Is there any evidence that the two processes are producing composite material with the same cross-sectional pattern to the fibres? If there is a difference, will it affect the specified properties of the composite? 

\item A model is proposed for a point pattern in the form of a simulation algorithm. There is no probabilistic definition of the model, although the algorithm is well-defined. An example of this would be a program which re-arranges a Poisson realization into a realization where each point is at the centre of a packed sphere [\cite{jt:1985}, \cite{lubstil:1990}]. Does the model fit the data? 

\end{enumerate}

The first of these examples is a standard problem in classification, but the other three would conventionally be seen as testing problems. Unfortunately, there are no classical tests which could be used in these examples.

\section{Tests for Spatial Point Patterns}

If a pattern is modeled by a Poisson process, then many tests are available to determine whether or not the process is Poisson [\cite{cressie}]. None of these are directly of any use in the last three problems. These tests are often based on a statistic or plot which can show signs of how the pattern is not Poisson in one particular way. Some of these statistics may be of use in summarizing aspects of non-Poisson point patterns. 

Formally expressed likelihood tests are not available for non-Poisson processes, on account of the lack of a formal and tractable expression for the joint distribution of the point locations. A test could be based upon the distribution of a descriptive statistic when the null hypothesis in the test was true. This requires a statistic whose null hypothesis distribution can be found. In general, there are no results beyond finding what the mean of the statistic should be. These results are few in number, and tend to be found in regular point processes with weak dependence among point locations within the realization. The pair correlation of the Matern Hard-Core point process can be explicitly found [\cite{stosto:1985}, \cite{stoy:1995a}]. This function could be used as a basis for a test based on estimates of the pair correlation function, provided that the variation of the estimates around the pair correlation function could be modeled. 

Simulation-based tests are of more use, yet still present difficulties. If a density is available for configurations of points (e.g. for Strauss processes), then likelihood ratio tests can be undertaken, based on MCMC methods [\cite{molwag}]. If no density is available, then a simulation test can be based on a statistic which is expected to detect important differences between the model and the data. This statistic is  calculated from many simulations of the null hypothesis model, and then compared with the value of the statistic found from the data [\cite{digg:1979}]. If the model and the data truly do differ, then finding that difference depends on the choice of the right statistic. In the comparison of a Poisson process with a Baddeley-Silverman cell process [\cite{badsil}], a comparison of $K-$functions would reveal no difference unless a Type I error occurred. Tests of this kind can be further complicated if the null hypothesis model requires that some parameters be estimated from the data.

\section{Summarization and Classification}

Assessing the fit of point process models to data using tests fails due to intractability or to the non-existence of useful probabilistic models. This approach also encounters  difficulties with finding a single statistic that can identify the differences that do exist. Often, there is no guidance at all beforehand as to which statistics will best identify differences.

In all three of the testing problems, one is trying to determine if there is any evidence of difference between two point processes. In the absence of a formal model for one or both of the processes, any information about differences has to come from looking at joint distributions of every statistic that could be calculated from their realizations. If the distributions of all possible finite collections of statistics are the same in both models, then one must conclude that the processes are indistinguishable. This can be called the \textit{strong inference} approach to model fitting, and is what physicists aspire to when they give up trying to show that a physical model (such as the quantum mechanical model for the hydrogen atom) doesn't fit. If differences are found, then it may be the case that these differences do not affect what the researcher is interested in. For example, if electrical conductance only depends on few other descriptive statistics of a sphere packing, then a model could be used if the joint distribution of electrical conductance and those statistics were the same for both processes. This approach  would allow simple models to be used on very complicated materials (such as concrete). It represents the engineering approach to finding the simplest usable model that fits the data faithfully in a particular case, rather than seeking a comprehensive theory. This approach can be called the \textit{weak inference} approach to model fitting for a specific property (electrical conductance, in this case), and it allows a lack of model identifiability to work in the scientist's favour.

In almost no cases is it possible to formally derive the distributions of the summarizing statistics, so it is necessary to estimate them from the data. This requires that any inference be based on large independent and identically distributed samples of realizations from both processes being compared. Single realization inference is not possible in the absence of a model. 

Given two large samples from both processes, statistical learning can be used as an exploratory tool to find evidence of differences between the two processes. The goal is to seek out statistics which best show the differences between the two sets of realizations, and then use these to build classification rules which distinguish single realizations from each process. If a classification rule [\cite{gordon}, \cite{htf}] can be constructed from a training set, this constitutes evidence that there could be a difference between the processes. If the classification rule continues to perform well on test sets of new realizations, then a low misclassification rate would be further evidence in favour of a difference and against the possibility that the classification rule was fit to noise. If the statistics required to identify a difference do not affect the distribution of a statistic of interest, then this could be taken to be evidence of a weak fit with respect to the statistic of interest. If no such classifier could be found after much effort, then this would suggest (in a manner aspiring towards strong inference) that the same process generates both sets of realizations. Since there is no end to the number of statistics that could be tried, establishing model fit by strong inference is impossible. The main use of the idea is to show how difficult it is to establish that a model fits comprehensively.

All of these methods are based on the analysis of data, and so there is always the risk that the samples of realizations will be unrepresentative, and that the differences found will be due to chance variation. The use of test sets and classifiers based on large samples of realizations reduces the chances of this happening, but cannot eliminate it. Insisting that the differences be so large as to give almost no misclassifications on large test sets further reduces the chance of this happening, although this would overlook differences in cases where the statistics found could only identify a minor but real difference in the distribution of statistics.

Using summarization and classification to build classification rules to distinguish between processes known to be different is also a good way to develop new and more powerful summarizing statistics. If there is a model as elegant and useful as the Gibbs or Poisson models for describing sphere packings, it may be based on statistics which have not yet been found. Experimental investigation may prove to be a more useful way of finding such models, rather than attempting to formally advance existing theory. 

In classification problems, this aggressive approach to finding differences using new statistics may be more effective than trying to build classification rules based on the classical statistics of point processes. 

\subsection{Summarization}

To make this approach to fitting work, more is needed than just access to effective classification rules. There must also be a very large library of descriptive statistics which can be applied to point patterns, many more than are currently in use now.  Distinguishing realizations of non-Poisson processes requires new statistics, which will need to be invented or borrowed from mathematical models used in other disciplines. 

\subsubsection{Point Process Statistics}

Most descriptive statistics for point patterns were developed in the context of inference for Poisson processes. The intensity, estimated for a homogeneous Poisson process as the number of points per unit area, describes the first moment properties of the underlying random measure. The second moment properties cannot be expressed by a single statistic, but can be described by the $K-$ function, the pair correlation function, and the nearest neighbour function. The last is related to the Palm distribution for the process, as is the empty space function. Third moment-based statistics can also be defined [\cite{schladitz}]. All standard estimators of these statistics are based upon the assumption that the point process is stationary. When the realizations come from a point process which is not stationary, then these statistics can still be used descriptively, as long as they are applied to each realization in a consistent fashion. 

\subsubsection{Random Set Statistics}

A realization of a point process can be transformed into a realization of a random set [\cite{mol:2005}] known as a germ grain model by attaching spheres at each point. If the process is regular with all points separated by distance $\delta$, then using discs of radius $\delta$ would be useful, although other radii could be used.   

The most basic random set statistics are estimates of the $k^{th}$ moments of the realization, which are defined by
\begin{equation*}
M_k(x_1, \ldots , x_k) = \text{Pr}[x_1, \ldots , x_k \in \Phi]
\end{equation*}
where $\Phi$ is the union of all of the spheres and $x_1, \dots , x_k \in \mathbb{R}^d$. When the point pattern is stationary, the reduced moments can be defined as
\begin{align}
m_1 &= \text{Pr}[0 \in \Phi]  \label{m1} \\
m_2(r) &= \text{Pr}[0, r \in \Phi]  \label{m2} \\
m_3(r,s) &= \text{Pr}[0, r, s \in \Phi], \label{m3}
\end{align}
where $0$ is the origin in $\mathbb{R}^d$.
Given a window of observation $A$ onto a realization of a stationary point process, the first moment can be estimated by finding the proportion of the area of A covered by spheres. Higher moments such as the third can be estimated by superimposing  translates of the realization by $r$ and $s$ onto the original and finding the proportion of the window intersection occupied by points where three spheres intersect.    
The third moment can be shown to be a smoothed function of the three point analogue of the pair correlation function [\cite{ts:1985}]. Since estimates of the third moment involve information used in estimating the second moment, some form of moment decomposition such as the third cumulant
\begin{equation*}
\kappa_3(r,s)=m_3(r,s)-m_1(m_2(r)+m_2(s)+m_2(r-s))+2m_1^3
\end{equation*}
should be estimated in order to emphasize those aspects of the third moment that are dependent on the use of all three defining points. 

\subsubsection{Tessellation-Based Statistics}

Given a realization of a point process, a Voronoi triangulation and its dual tessellation [\cite{obs}] can be constructed. Statistics based on tessellations and triangulations are often used in the study of point processes arising from sphere packings [\cite{finney}, \cite{jm:2000}, \cite{aste}]. In the case of regular point processes, the tessellation will provide a skeleton for the void structure around the points. In general point patterns, the near neighbours of a given point can be defined as all points linked to the given point by triangulation edges. The triangulation is a slightly more useful source of statistics, since each cell has a fixed shape. 

Given a triangulation or a tessellation, various statistics can be calculated directly. These include means, standard deviations, maxima, and minima of cell volume, cell surface area, cell perimeter, longest side, and largest angle. For tessellations, the mean number of sides is also useful.  

Other aspects of tessellations can also be used as statistics. Given any listing of the $n$ points in a realization, the adjacency matrix $A$ is a sparse $n \times n$ matrix whose non-zero-entries $A_{ij}=1$ iff points $i$ and $j$ are connected by an edge of the triangulation. The three largest eigenvalues of this matrix, as well as the coefficients of its characteristic polynomial can be used as descriptive statistics. The coefficients of the characteristic polynomial include the number of triangles in the triangulation [\cite{biggs}], which can be used as a statistic when the counts from all realizations are adjusted for any scale or window size differences. 

\subsubsection{Statistics Based on Physical Models}

Germ-grain models based on regular point processes can be used as models  for composite materials [\cite{torq}],  for the liquid state [\cite{altil}], and for granular materials [\cite{thorntony}]. Physicists generally want to evaluate some bulk property on each individual realization, such as viscosity (in liquids), bulk electrical conductance (in composites), or yield stress (in granular materials). The mathematical models and algorithms for finding these properties from realizations can be used out of their physical context in order to provide statistics [\cite{picka:2007a}]. Simplifications of the structure of a realization using tessellations and triangulations are often needed to make these statistics calculable. 

\subsubsection{Smoothly Defined and Structurally Defined Statistics}

Summarizing statistics can be divided into two classes, based on how they extract information from a realization of point process. In both cases, some form of measurement is taken at many different locations in the observation. A [\textit{smoothly defined}] statistic reduces the list of measurements to a single number through use of some function that is invariant to the ordering of the measurements. One example is the volume fraction of a realization, which can be estimated as the average of a set of indicator function values on a grid of points within the realization. All moment estimates, all histograms of measurements of this kind, and generally all statistics based on the order statistic of the measurements are smoothly defined.  

A \textit{structurally defined} statistic is one whose treatment of the measurements is determined by large scale random structures within the realization. If a triangulation is assumed to be electrically conducting with edge resistances proportional to their length, then the bulk resistance of that circuit is proportional to a weighted average of lengths of loop-free paths between the electrodes in the circuit.  Bulk resistance can be used to define a measure of anisotropy for sphere packings [\cite{picka:2005}]. The weights are determined by a physical model for electrical conduction, but the paths are determined by the overall structure of the entire realization. 

Smoothly defined statistics make up most of the statistics currently used to summarize point patterns and random sets. They are of great use for describing Poisson process realizations, but they are difficult to interpret when applied to regular point processes. Structurally averaging statistics are expected to be of great use in the study of regular point processes, since their form is determined by the very disorder whose detail is smoothed away by the smoothly defined statistics.

\subsection{Classification}

To build a classification rule, a large collection of labeled realizations from both processes is needed for use as a training set. 
If the right summarizing statistics have been found and the realizations are sufficiently large, then the vectors of $k$ summarizing statistics for each element in the training set should form two distinct clouds of points in $\mathbb{R}^k$. The goal of classification is to find a hyperplane or hypersurface which separates the two clouds of points. Ideally, the hypersurface will completely separate the two clouds of points.

It is impossible to clearly visualize the two clouds of points that the classification rule separates if the number of statistics exceeds 3. Theory can provide some idea of the shape of the clouds, provided that the statistics involved are smoothly defined averages. Then, even in the case of a point pattern arising from a sphere packing, there are results which suggest that the between-realization distribution of those statistics will be Normal [\cite{penyu}]. This suggests that in those circumstances, the clouds would roughly ellipsoidal and that linear or quadratic discrimination should be effective. If some of the statistics are not smoothly defined averages and are instead based on extrema, then the limit laws may be quite different and the shape of the point clouds would be unclear. In these circumstances, non-linear methods such as $k-$nearest-neighbours, CART, or boosted or bagged CART would be required to separate the two clouds. 

\section{Examples}

Classification rules were developed to separate realizations of point processes in which the processes were known to be different. Three pairs of processes were considered. In each pair, one process was a hard-core Strauss process [\cite{strauss},  \cite{molwag}, \cite{geyer:1999}] chosen to have an intensity close to that of the other process in the pair. The processes compared to the Strauss Processes were the Dead Leaves (DL) process [\cite{serr:1982}], the Simple Sequential Inhibition (SSI) process [\cite{dbg}], and the Diggle Gratton (DG) process [\cite{digrat}] with parameters $\kappa = 3$, $\delta = 2$, and  $\rho=1$. All processes were simulated on a 44 by 44 unit window with the central 39 by 39 unit portion used as the realization. No points were allowed to be closer than 2 units. In the Diggle-Gratton process, there was an additional weak exclusion effect in a region beyond the unit hard-core exclusion zone around each point. When discs of unit radius are attached to each point, the mean estimated  coverage fractions are 0.24 for the Dead Leaves model, 0.39 for the SSI model, and 0.51 for the Diggle-Gratton model.

\begin{figure}[htbp] 
       \centering
       \includegraphics[bb= 18 18 577 824, angle = 270, width=5in]{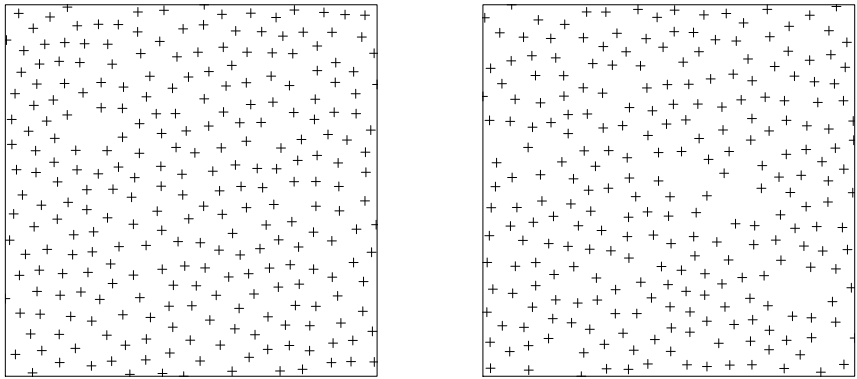}  
       \caption{Typical realizations (Diggle-Gratton at left, Strauss at right)}
       \label{discern}
    \end{figure}

The training sets for each pair of processes consisted of 300 realizations of each kind. The testing sets consisted of 100 realizations of each kind.

The nearest neighbour functions $G(r)$ were estimated for each realization in spatstat [\cite{bt:2005}] and the distances from all training set realizations were combined. The model
\begin{equation*}
\log(1-G(r))= \alpha (r-2) + \beta L(r-2) + \epsilon
\end{equation*}
where $L(x)=(3x^2-1)/2$ and $\epsilon \sim N(0,\sigma)$ was found to fit these distances well, and so the model was fit to each realization. The fit was quadratic without intercept, and the quadratic term used was the second-order Legendre polynomial. The estimates of $\alpha$, $\beta$, and $\sigma$ for each realization were used as descriptive statistics.

The remaining summarizing statistics were found from the Delaunay triangulation of the points, as determined by spatstat. Once the triangulation had been cleaned of artificial vertices and edges, the number of triangles was determined. The longest side, largest angle, and area were found for each triangle, and the maxima of each were determined for each realization. The three largest eigenvalues of the adjacency matrix derived from the triangulation were also used as statistics.

\begin{table}[htdp]
\caption{Misclassification rates (MR) and CIs based on all 10 statistics}
\begin{center}
\begin{tabular}{|l|cc|cc|}\hline
Case & MR Strauss  & MR Other  & 95\% CI MR Strauss & 95\% CI MR Other\\
  \hline
DL & 18/100 & 18/100 & (0.110, 0.270) & (0.110, 0.270) \\
SSI & 13/100 & 21/100 & (0.071, 0.210) & (0.130, 0.303) \\
DG & 0/100 & 0/100 & (0, 0.03) & (0, 0.03) \\ \hline
\end{tabular}
\end{center}
\label{t1}
\end{table}

Misclassification rates for linear discriminant analysis are given in Table \ref{t1}. The summarizing statistics were not useful for distinguishing the Dead Leaves and SSI processes from the corresponding Hard-Core Strauss process, although the classification rules did slightly better than guessing. The Diggle-Gratton rule made no classification errors. Performance on the rule on the training sets were statistically indistinguishable from their performance on the testing sets. 

\begin{table}[htdp] \caption{Misclassification rates (MR) and CIs for two non-Strauss training sets from the same model}
\begin{center}
\begin{tabular}{|l|cc|cc|}\hline
Case & MR [Sub 1]  & MR [Sub 2]  & 95\% CI MR [Sub 1] & 95\% CI MR [Sub 2]\\
  \hline
DL & 19/50 & 23/50 & (0.246, 0.523) & (0.318, 0.607) \\
SSI & 22/50 & 27/50 & (0.200, 0.587) & (0.393, 0.682) \\
DG & 25/50 & 23/50 & (0.355, 0.645) & (0.318, 0.607) \\ \hline
\end{tabular}
\end{center}
\label{t2}
\end{table}

Classification rules were also found for each of the non-Strauss data sets, based on a division of those data sets into two arbitrary halves. The results in Table \ref{t2} for the non-Strauss processes suggests that the classification rule has a quality analogous to good Type I error performance. 

\begin{figure}[htbp] 
       \centering
       \includegraphics[bb= 18 18 577 824, angle= 270, width=5in]{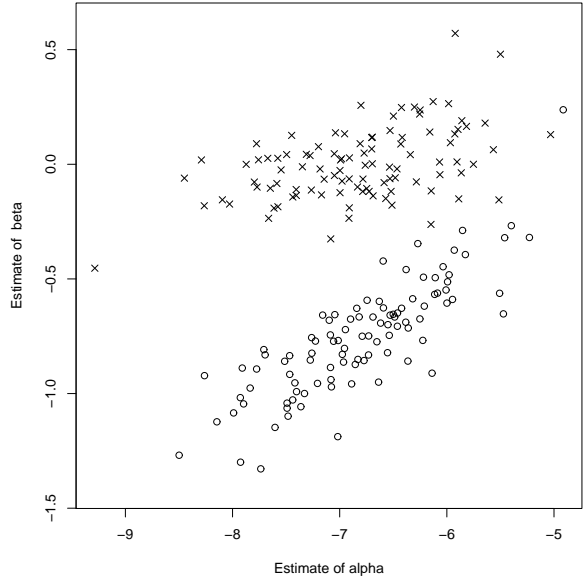}  
       \caption{Unpooled Diggle-Gratton ($\circ$) and Strauss Process ($\times$) estimates of $\alpha$ and $\beta$. }
       \label{g2}
    \end{figure}

The dominant statistics in the classification were the three arising from the nearest-neighbour function. Using only these statistics slightly raised the number of misclassifications in each case or left it unchanged. The estimates of $\alpha$ and $\beta$ for the Diggle-Gratton case are shown in Fig. \ref{g2}.

\begin{figure}[htbp] 
       \centering
       \includegraphics[bb= 18 18 577 824, angle = 270, width=4.5in]{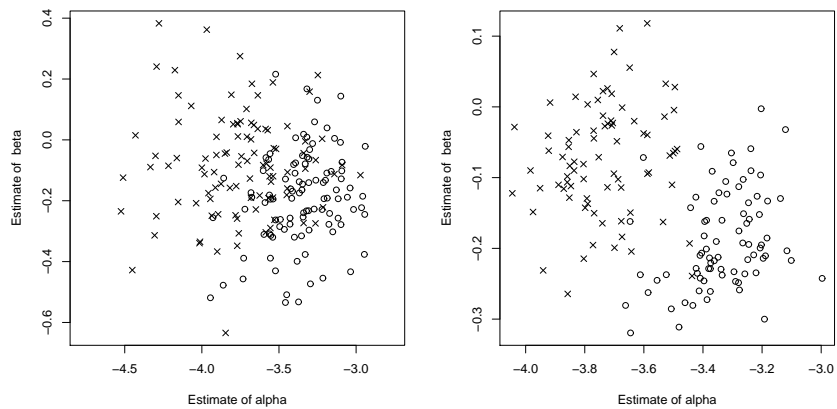}  
       \caption{SSI ($\circ$) and Strauss Process ($\times$) estimates of $\alpha$ and $\beta$ for unpooled data (left) and pooled data (right). }
       \label{ssi}
    \end{figure}

To investigate the effect of using more data, the 400 realizations from each process were pooled into 80 sets of 5 realizations each. This would correspond to a situation in which multiple different samples are taken at widely spaced locations from the same specimen of material. Of the 80 realizations, 60 were used in training sets and 20 were used for testing. The triangulation and eigenvalue statistics were combined by taking the minimum value of each, out of the 5 realizations. The nearest neighbour statistics were found by pooling the nearest-neighbour distances from all 5 realizations and fitting the same model as in the previous case. Table \ref{t3} suggests that this could lead to better classifier performance, but a better way to see the effect is to consider the degree to which pooling separates the estimates of $\alpha$ and $\beta$ for the SSI case (Fig. \ref{ssi}). The improvement in separation is more clearly visible than through the misclassification rate, although it is based on fewer data points in the pooled case. 

\begin{table}[htdp]
\caption{Misclassification rates and CIs for pooled case}
\begin{center}
\begin{tabular}{|l|cc|cc|}\hline
Case &  MR Strauss  & MR Other  & 95\% CI MR Strauss & 95\% CI MR Other\\
  \hline
DL & 1/20 & 0/20 & (0.001, 0.250) & (0, 0.139) \\
SSI & 1/20 & 2/20 & (0.001, 0.250) & (0.012, 0.320) \\
DG & 0/20 & 0/20 & (0, 0.139) & (0, 0.139) \\ \hline
\end{tabular}
\end{center}
 \label{t3}
\end{table}

In all cases, linear discriminant analysis produced the lowest misclassification rates. Logistic discrimination performed about as well, but CART and 5-Nearest-Neighbours produced rules which often did no better than guessing. This was attributed to the relatively small size of the training set.


The example shows that summarization and classification is useful way to build classifiers. It does depend very heavily on having powerful summarizing statistics, which were available only in the Diggle-Gratton case. In the other cases, further experimentation is needed to see if an effective rule can be found based on larger realizations, or if more powerful statistics are needed. The classification rule  in the Diggle-Gratton case was not based on many more nearest-neighbour distances than was the rule for the SSI case. Extra classification power may have been derived from the Diggle-Gratton case data having an intensity closer to that of sphere packings, which can be shown to be less disordered than other regular point processes [\cite{picka:2007b}]. As with all conclusions that are based solely on data, their strength can only be determined through analysis of more realizations. 

If the investigator is clever enough to find the right summarizing statistics and has enough data, then very powerful classifiers can be built. While the statistics are different than those used by the eye and mind of a trained pathologist (in the case of pathological specimens), the fact that a classification rule can be built which can correctly classify 200 realizations of the kinds seen in Fig. \ref{discern} suggests that a rule built by summarization and classification could do at least as well.


\end{document}